\documentclass[12pt]{iopart}
\usepackage{epsfig}

\begin{document}
\pagestyle{empty}
\vspace*{-1.8cm}
\begin{flushright}{\bf LAL 01-70}\\
{October 2001}\\
\end{flushright}
\vskip 1.5 cm

\begin{center}
{\bf\LARGE Structure functions at large $x$} 
\end{center}
\vspace*{1.5cm}
\begin{center}
{\bf\Large Zhiqing Zhang}\\
(On behalf of the H1 and ZEUS collaborations)

\vspace*{0.5cm}

\normalsize {    {\bf 
Laboratoire de l'Acc\'el\'erateur Lin\'eaire}}\\
\small {    {
CNRS-IN2P3 et Universit\'e de Paris-Sud - B\^at. 200 - BP 34 - 91898 Orsay cedex}}\\
\end{center}
\vskip 2.5truecm

\begin{abstract}
Structure function data together with other measurements from fixed-target 
deep inelastic scattering
and hadron-hadron collider experiments which contribute to our knowledge of 
the parton density functions are reviewed. The inclusive cross-section
measurements of neutral and charged current interactions at HERA are
presented and their impact on the parton density functions is discussed.
Future prospects for an improved knowledge of the parton density functions
at large $x$ are briefly mentioned.
\end{abstract}
\vskip 4.5truecm

\begin{center}
{\it Invited talk given at the workshop on ``New Trends in HERA 
Physics 2001'',\linebreak Ringberg castle, Germany, June 17-22, 2001.}
\end{center}
%Uncomment for PACS numbers title message
%\pacs{00.00, 20.00, 42.10}

% Uncomment for Submitted to journal title message
%\submitto{\JPA}

% Comment out if separate title page not required
\maketitle
\newpage
\quad
\newpage
\pagestyle{plain}
\section{Introduction}
Deep inelastic scattering (DIS) experiments have played important roles in
understanding the partonic structure of hadrons and in establishing the theory 
of quantum chromodynamics (QCD), the strong sector of the Standard Model (SM).
Our current knowledge of the parton densities in hadrons is primarily
derived from the structure functions measured in these
experiments~\cite{zhang}.

A precise knowledge of these parton density functions (PDFs) is needed
both for providing reliable predictions for processes involved in
hadron-hadron colliders such as the Tevatron and the LHC, and for achieving
precision measurements at these colliders. A good example is the measurement
of the mass of the $W$ boson from the Tevatron. The mass, which is relevant
for incisive tests of the SM of electroweak interactions, receives 
a non-negligible systematic contribution from the uncertainty of 
the PDFs~\cite{mwcdfd0}.

The precision of the PDFs also affects directly interpretations of data
measured at hadron colliders and searches for physics beyond the SM.
One example is the excess of jet events at large transverse
energies over perturbative QCD calculations reported earlier by the CDF
collaboration~\cite{cdfjet}.
The excess, which has initiated considerable speculations of possible
new physics, could well be accommodated by a higher than expected
gluon density at large $x$~\cite{huston}. Another example is the excess of
events at high momentum transfer ($Q^2$) with respect to the standard DIS
expectation reported by both the H1 and ZEUS experiments at HERA
based on the earlier low statistics sample taken from 1994 to
1996~\cite{h1zeushq97}.
The excess could be due to a statistical fluctuation, an imprecise knowledge 
of the PDFs at large $x$, or a resonance production of leptoquarks or
scalar quarks in $R_p$-violating supersymmetric models~\cite{altarelli97}.

In this paper, various constraints on the PDFs from fixed-target DIS 
experiments and from hadron-hadron colliders will first be briefly reviewed 
(section~\ref{sec:constraints}), the inclusive cross-section measurements 
at high $Q^2$ from HERA will then be presented and their impact on the PDFs 
discussed (section~\ref{sec:hera}).

\section{Current knowledge on the parton density functions}
\label{sec:constraints}
\begin{figure}
\begin{center}
\begin{picture}(50,140)
\put(-120,-10)
{\hspace*{2cm}\epsfig{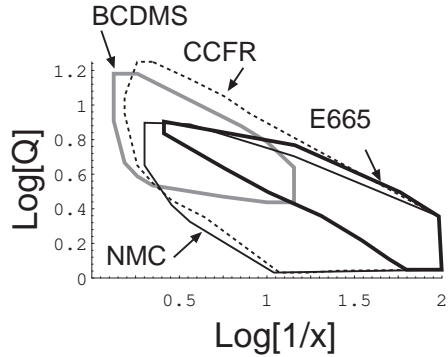}}
\end{picture}
\end{center}
\caption{\label{fig:ftdis}The kinematic coverage~\cite{albrow97} 
of four precise
structure function measurements by BCDMS~\cite{bcdms}, CCFR~\cite{ccfr}, 
E665~\cite{e665} and NMC~\cite{nmc}.}
\end{figure}

The traditional, but still the most important, constraint on the PDFs is from
structure function data measured in DIS experiments.
Shown in figure~\ref{fig:ftdis} are four precise measurements from 
BCDMS~\cite{bcdms}, CCFR~\cite{ccfr}, E665~\cite{e665} and  NMC~\cite{nmc} 
and their kinematic ranges~\cite{albrow97}.
These data constrain the PDFs mainly in the medium and large $x$
range\footnote{The best knowledge on the PDFs at small $x$ is obtained
from HERA structure function $F_2$ data. This is, however, not the subject of 
this paper.}.

The precision of the structure function data does not, however, imply a good
precision for the PDFs since their derivation and error estimation depend
on a whole complexity of experimental and theoretical inputs involved in a
global analysis such as MRST~\cite{mrst} or CTEQ~\cite{cteq}. Here are a
number of sources of uncertainty:
\begin{itemize}
\item {\bf Experimental systematic uncertainties:} The most precise data are
usually limited by systematic, rather than statistical, uncertainties.
The correlations of different systematics on the measurements among 
and across experiments are not at all trivial to take properly into
account.
\item {\bf Higher-twist contribution:} The data are located at relative low
$Q^2$ and large $x$. The higher-twist contribution is expected to be
important as it behaves as $[(1-x)Q^2]^{-1}$ with respect to the
leading-twist contribution.
\item {\bf Parameterization form:} In a global analysis, the parton
densities are parameterized in a certain functional form. The freedom in
choosing the functional form and the corresponding initial scale is a source
of the uncertainty.
\item {\bf Large nuclear corrections:} As far as the $d$ valence quark
density is concerned, it is constrained mainly by the deuterium data, to which
the nuclear binding corrections can be important.
\end{itemize}

\begin{figure}[h]
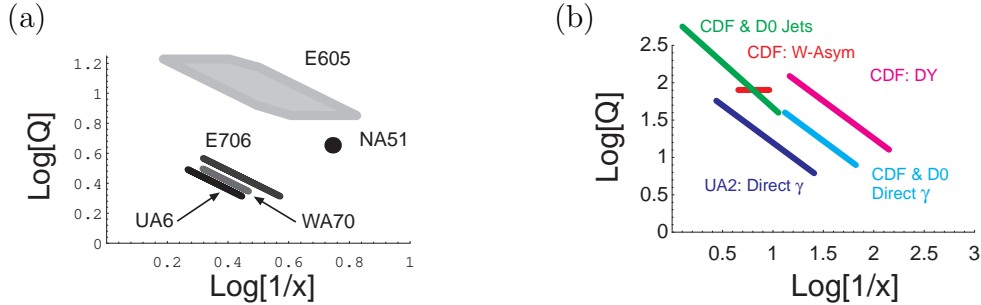

\vspace*{-1.2cm}
\begin{center}
\begin{picture}(50,140)
\put(-165,95){(a)}
\put(-130,-10){\hspace*{-1cm}\epsfig{figure=zhan.dph.eps,bbllx=0pt,bblly=0pt,bburx=594pt,
bbury=842pt,width=13cm}}
\put(42,95){(b)}
\put(85,-10){\hspace*{-1cm}\epsfig{figure=zhan.hhc.eps,bbllx=0pt,bblly=0pt,bburx=594pt,
bbury=842pt,width=11.cm}}
\end{picture}
\end{center}
\caption{\label{fig:dph}The kinematic coverage~\cite{albrow97} of 
(a) the lepton-pair
(Drell-Yan) production from E605~\cite{e605}, the asymmetry data in 
the Drell-Yan production from NA51~\cite{na51}, and the direct photon 
production data from E706~\cite{e706}, WA70~\cite{wa70} and UA6~\cite{ua6} 
and of (b) the direct photon ($\gamma$) data from UA2~\cite{ua2},
CDF~\cite{cdfdph} and D0~\cite{d0dph}, the Drell-Yan (DY) and 
$W$ asymmetry data from CDF~\cite{cdfdy,cdfwasy}, and the jet data from
CDF~\cite{cdfjet} and D0~\cite{d0jet}. 
The kinematic ranges of some of the
measurements are now extended by new experiments: e.g.
E605~\cite{e605} by E772~\cite{e772}, NA51~\cite{na51} by E866~\cite{e866}.}
\end{figure}

Apart from the structure function data, a few other processes from 
the fixed-target experiments also provide valuable constraints on the PDFs.
This is the case for the lepton-pair production or the Drell-Yan process
(the dominant leading-order subprocess being $q\overline{q}\rightarrow
\gamma^\ast g, \gamma^\ast\rightarrow l\overline{l}$),
which constrains the sea quark distributions in the proton. The asymmetry
between the $u$ and $d$ quark flavors, which cannot be easily determined
from the structure function data, is constrained by the asymmetry in the
Drell-Yan production.
The direct constraint on the gluon density could in principle be obtained from
the prompt photon production ($qg\rightarrow \gamma X$). However, 
the large discrepancies between measurements and theoretical predictions and 
among measurements carried out by different experiment groups~\cite{laenen}
prevent us from using these data at present. 
Represented in figure~\ref{fig:dph}(a) are the kinematic ranges for these
processes. Figure~\ref{fig:dph}(b) shows other
constraining processes from the collider experiments UA2, CDF and D0. 
In addition to
the direct photon process, the $W$ asymmetry constrains the $d$ over $u$
ratio, $d/u$, for $x$ around 0.1 and for $Q$ around the mass of the $W$ boson.
The inclusive jet data from the Tevatron provide a potential source for 
constraining the gluon density at large $x$.

\begin{figure}[t]
\vspace*{-6.5cm}
\begin{center}
%\begin{picture}(50,360)
%\put(-130,-20){
\epsfig{figure=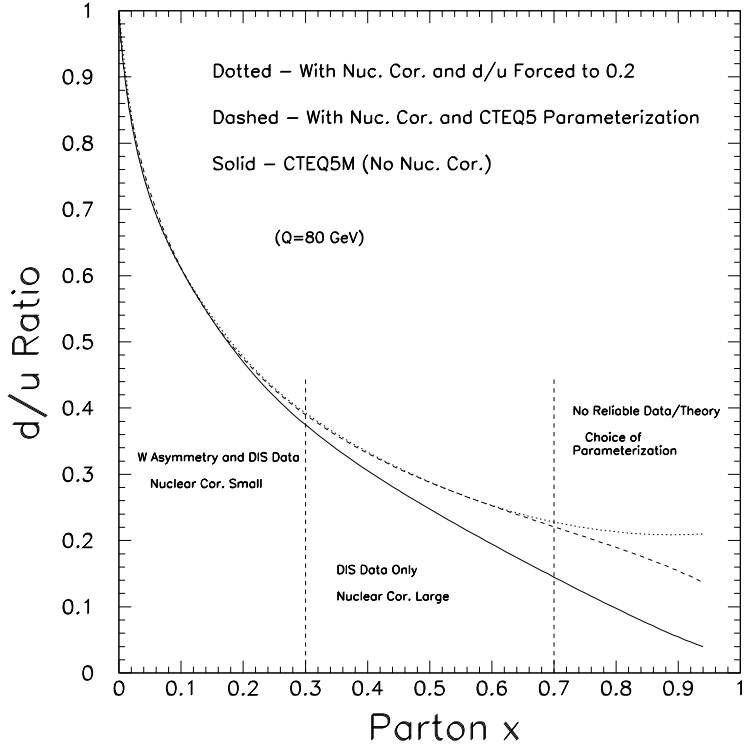,bbllx=0pt,bblly=0pt,bburx=594pt,
bbury=842pt,width=11.5cm}
%}
%\end{picture}
\end{center}
\vspace*{-0.5cm}
\caption{\label{fig:dou}The $d/u$ ratio as a function of $x$ for
$Q=80$\,GeV~\cite{kuhlmann}.
The solid (dashed) curve corresponds to the parameterization from CTEQ5M
without (with) nuclear corrections applied. The dotted curve shows the result
when the nuclear corrections are applied and the ratio is forced to be 0.2 at
$x=1$.}
\end{figure}

Despite the large kinematic coverage of the different data, the resulting
uncertainty of various PDFs is far from uniform. In general, the precision is
best in the medium region but still rather poor towards the kinematic limit
at large $x$. This is well illustrated by the behavior of the $d/u$ ratio at
$x\rightarrow 1$ (figure~\ref{fig:dou}). On the theoretical side, the model predictions vary
considerably between 0 and 0.5 with non-perturbative QCD-motivated predictions
at around 0.2~\cite{farrar75}.
On the experimental side, for $x<0.3$, there are precise data from both
$W$ asymmetry and DIS data and the nuclear corrections to the DIS data are
insignificant; for $0.3<x<0.7$, there are only DIS data which may be subject
to large nuclear corrections; at larger $x$, no reliable data are available.
In an analysis by Yang and Bodek~\cite{yang}, they showed that the description
of the $W$ asymmetry and NMC structure function ratio data is improved with
$d/u\rightarrow 0.2$
as $x\rightarrow 1$ and with the  nuclear binding corrections applied to
the deuterium data. The analysis by Kuhlmann \etal~\cite{kuhlmann} showed
that if the ratio $d/u$ is around 0.2,
the NMC data indeed need a nuclear correction, but the converse is
not necessarily true.
The large spread of the curves shown in figure~\ref{fig:dou}, corresponding
to three possible fits to the existing data, shows how uncertain the
current PDFs are at large $x$.

\section{HERA impact on the parton density functions} \label{sec:hera}
The structure functions measured by the HERA experiments H1 and ZEUS have
provided a unique constraint on the PDFs at small $x$, in particular 
on the gluon density~\cite{zhang}. 
Here we shall present inclusive cross-sections at high
$Q^2$ measured with three important data samples collected by both
experiments since 1994, and discuss their impact on the PDFs at large $x$.
The first $e^+p$ data sample, corresponding to an integrated
luminosity of 35.6\,pb$^{-1}$\footnote{The number shows the
integrated luminosity from H1. The data samples from ZEUS are comparable.},
was taken from 1994 to 1997 at a center-of-mass energy of 300\,GeV. Both
the $e^-p$ data of 1998-1999 and the $e^+p$ data of 1999-2000 are taken at a
center-of-mass energy of 320\,GeV resulting from the increased proton energy
of 920\,GeV. The corresponding integrated luminosities are, respectively,
16.4\,pb$^{-1}$ and 65\,pb$^{-1}$.

\begin{figure}[h!]
\vspace*{-1.8cm}
\begin{center}
%\begin{picture}(50,500)
%\put(-140,-30){
\epsfig{figure=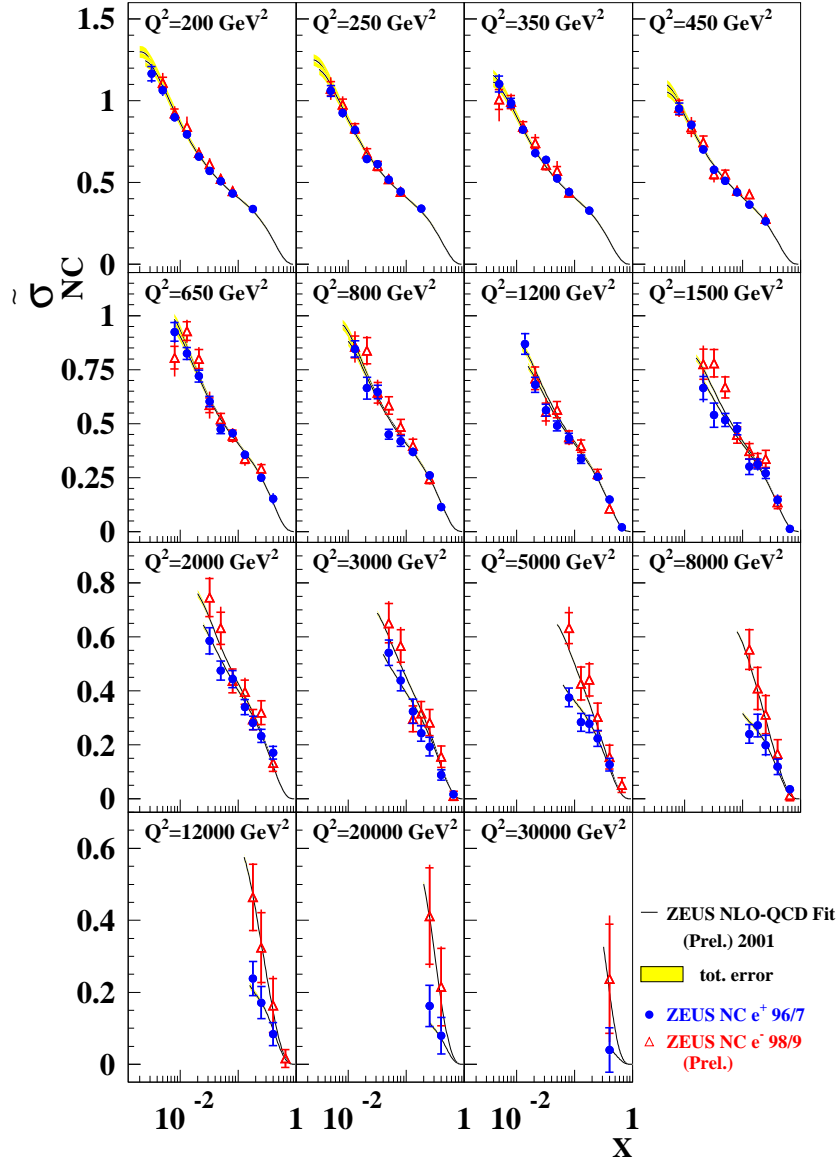,bbllx=0pt,bblly=0pt,bburx=594pt,
bbury=842pt,width=12.5cm}
%}
%\end{picture}
\end{center}
\vspace*{-0.5cm}
\caption{\label{fig:nczeus}The NC reduced cross-sections measured by ZEUS
using the 1996-1997 $e^+p$ and 1998-1999 $e^-p$ data~\cite{zeus9697,zeus9899}.
The curves are the
preliminary ZEUS next-to-leading (NLO) order fit~\cite{zeusfit} based on 
the fixed-target data and the 1996-1997 ZEUS data.}
\end{figure}

\begin{figure}
\vspace*{-5cm}
\begin{center}
%\begin{picture}(50,335)
%\put(-140,-30){
\epsfig{figure=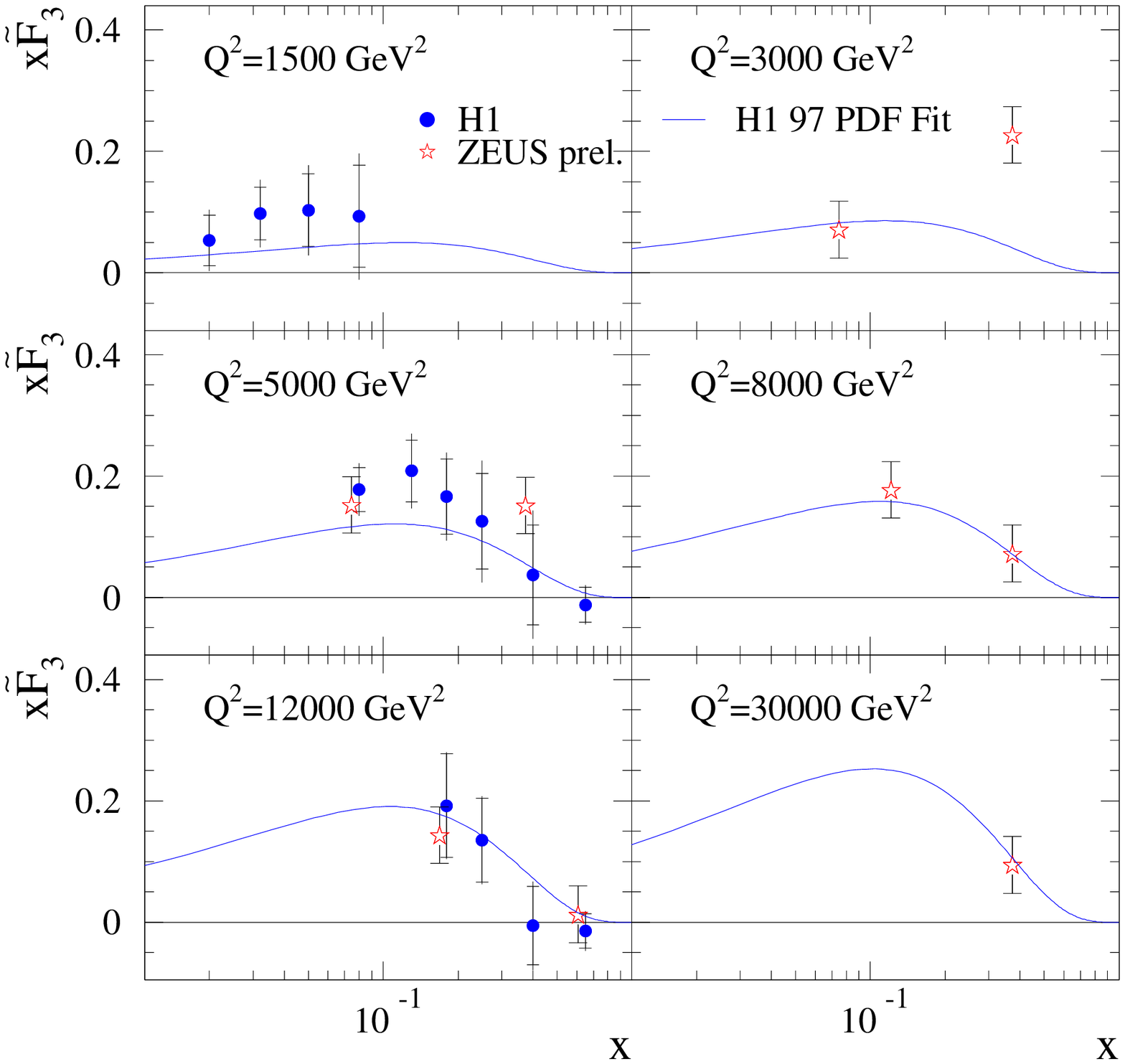,bbllx=0pt,bblly=0pt,bburx=594pt,
bbury=842pt,width=10.5cm}
%}
%\end{picture}
\vspace*{-1.cm}
\end{center}
\caption{\label{fig:xf3}The structure function $x\tilde{F}_3$ measured by H1
and ZEUS~\cite{h19899,zeus9899}. The curves are the results of 
the H1 97 PDF Fit~\cite{h19497}.}
%\end{figure}
%
%\begin{figure}[h!]
\vspace*{-3cm}
\begin{center}
%\begin{picture}(50,335)
%\put(-140,-115){
\epsfig{figure=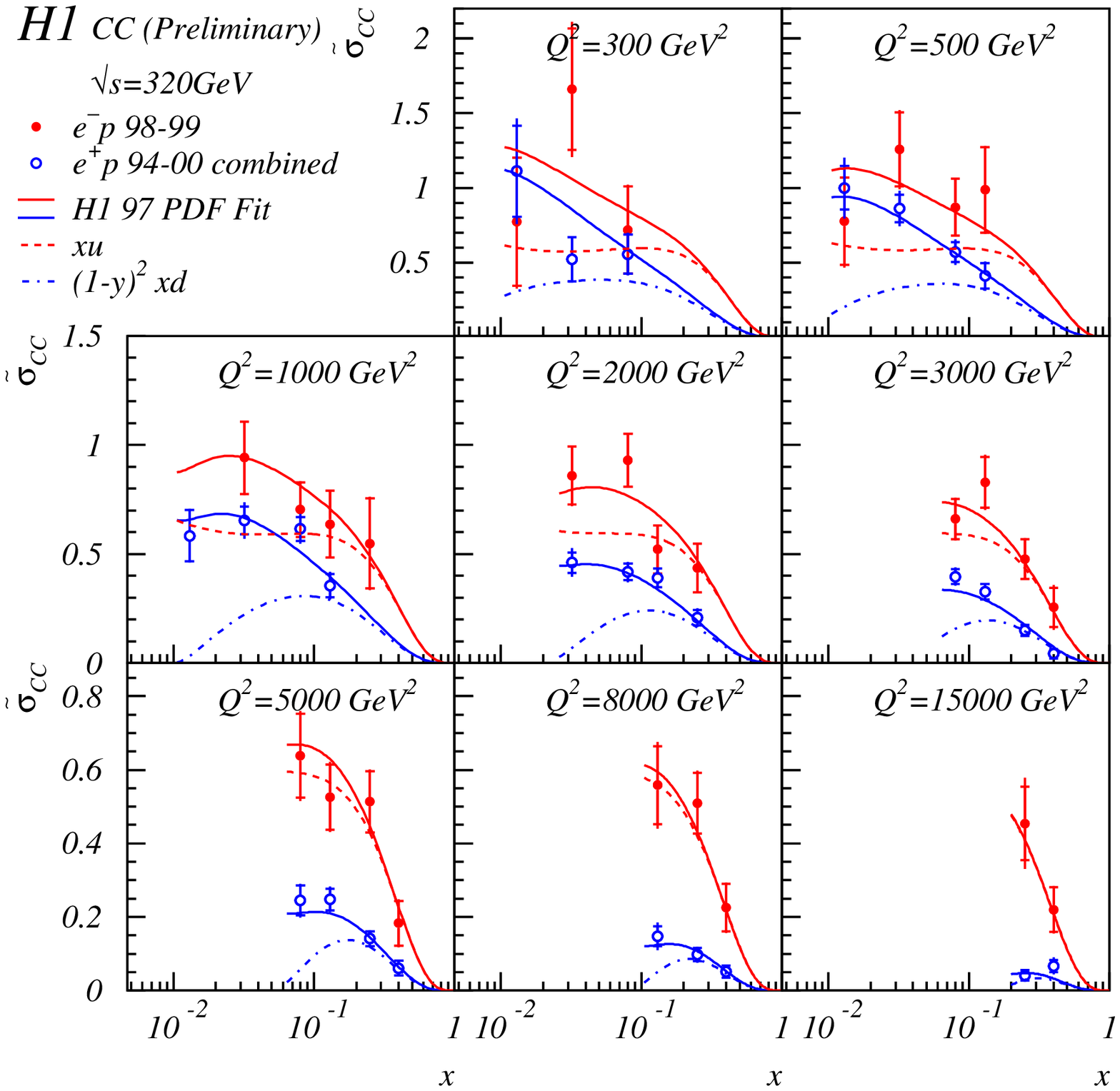,bbllx=0pt,bblly=0pt,bburx=594pt,
bbury=842pt,width=10.5cm}
%}
%\end{picture}
\end{center}
\vspace*{-3cm}
\caption{\label{fig:cch1}The CC reduced cross-sections $\tilde{\sigma}_{\rm
CC}$ measured by H1 using the 1998-1999 $e^-p$ and combined 1994-1997 and 
1999-2000 $e^+p$ data~\cite{h19497,h19899}.
The solid curves show the expectations based the H1 97 PDF Fit~\cite{h19497}. 
The dashed and dash-dotted curves represent, respectively, the contribution of 
$xu$ and $(1-y)^2xd$ to the $e^-p$ and $e^+p$ cross-sections.}
\end{figure}

Figure~\ref{fig:nczeus} shows the neutral current (NC) reduced 
cross-sections\footnote{The NC reduced cross-section $\tilde{\sigma}_{\rm NC}$ 
is defined as $\tilde{\sigma}=(xQ^4/2\pi\alpha^2Y_+)\rmd^2\sigma/\rmd x\rmd 
Q^2$ with $Y_+=1+(1-y)^2$.} measured with the 1996-1997 $e^+p$ and
1998-1999 $e^-p$ ZEUS data~\cite{zeus9697,zeus9899}. The $e^+p$ and $e^-p$ cross-sections are found to be comparable\footnote{The
difference due to the change in the center-of-mass energies is expected at a
few percent level.} at low $Q^2$ ($<1000\,{\rm GeV}^2$ or so).
This is understood to be due to the dominance of $\gamma$ exchange. 
At higher $Q^2$,
the $e^-p$ cross-sections are measured to be increasingly larger than
those of $e^+p$, demonstrating the $\gamma-Z$ interference contribution.
The cross-section asymmetry allows the structure function $x\tilde{F}_3$ 
(figure~\ref{fig:xf3}) to be determined for the first time at
HERA~\cite{zeus9899,h19899}. As this structure function measures
the difference between the quark and anti-quark densities ($x\tilde{F}_3\sim
2\sum_ie_ia_ix(q_i-\overline{q}_i)$ with $e_i$ and $a_i$ being, respectively,
the electric charge of quark $i$ and its axial coupling to the $Z$ boson),
it is thus sensitive to the valence quark densities at large $x$. 
\newpage
\quad

A comparison of the charged current (CC) reduced cross-sections\footnote{The 
CC reduced cross-section $\tilde{\sigma}$ is defined as $\tilde{\sigma}=(2\pi
x/G^2_F)((Q^2+M^2_W)/M^2_W)^2\rmd^2\sigma/\rmd x\rmd Q^2$ with $G_F$ and
$M_W$ being, respectively, the Fermi coupling constant and the mass of the
exchanged $W$ boson.} measured by H1~\cite{h19497,h19899} is shown in
figure~\ref{fig:cch1}. The difference in the cross-sections results mainly
from different quark flavors probed by the exchanged $W^\pm$ bosons. 
The CC cross-sections at high $Q^2$ thus provide a unique source to 
directly constrain the $u$ and $d$ valence quarks at large $x$.
%\newpage

\begin{figure}[h!]
\vspace*{-1.5cm}
\begin{center}
\begin{picture}(50,400)
\put(-155,-105){\epsfig{figure=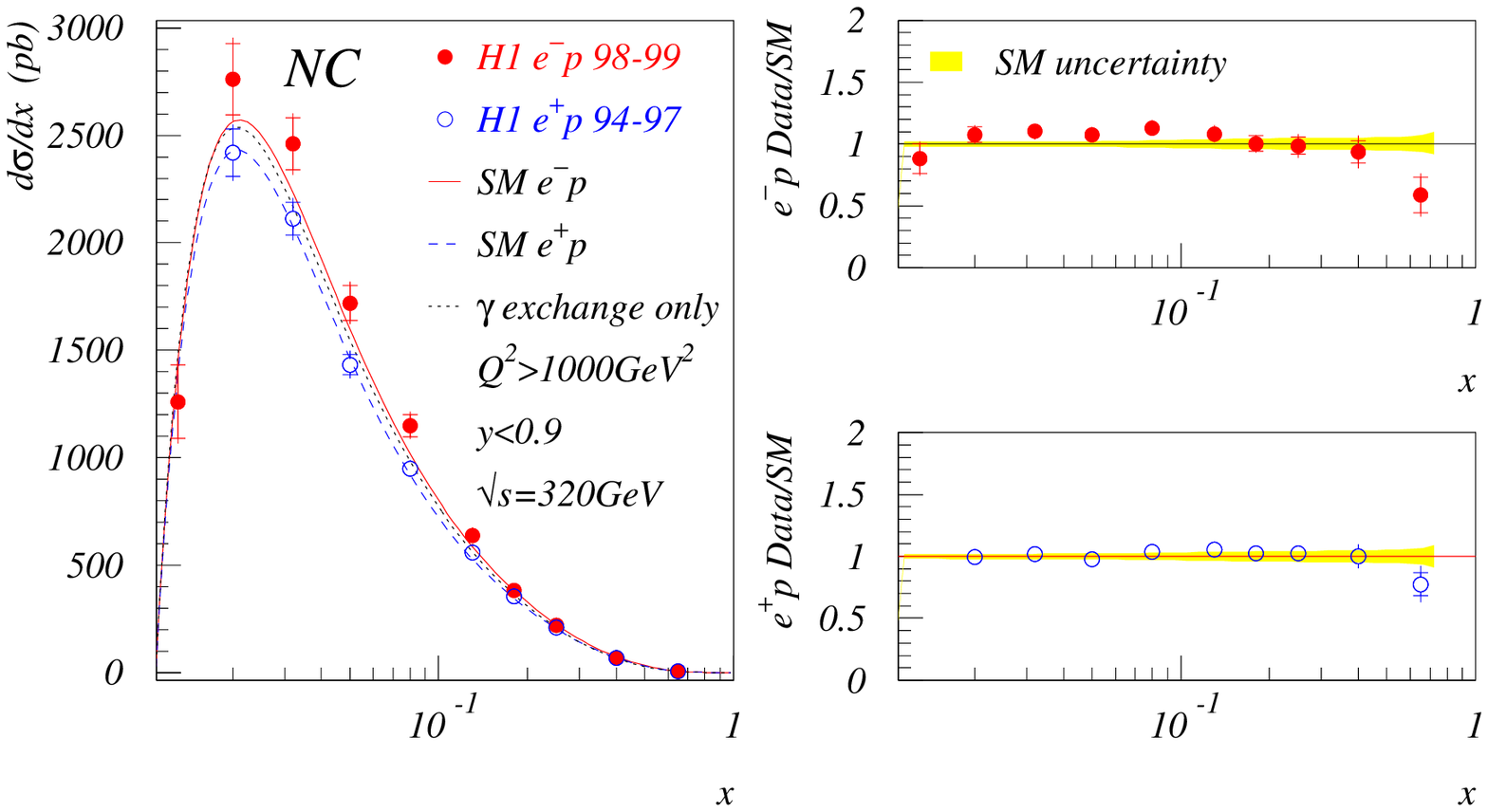,bbllx=0pt,bblly=0pt,bburx=594pt,
bbury=842pt,width=14cm}}
\put(-155,-295){\epsfig{figure=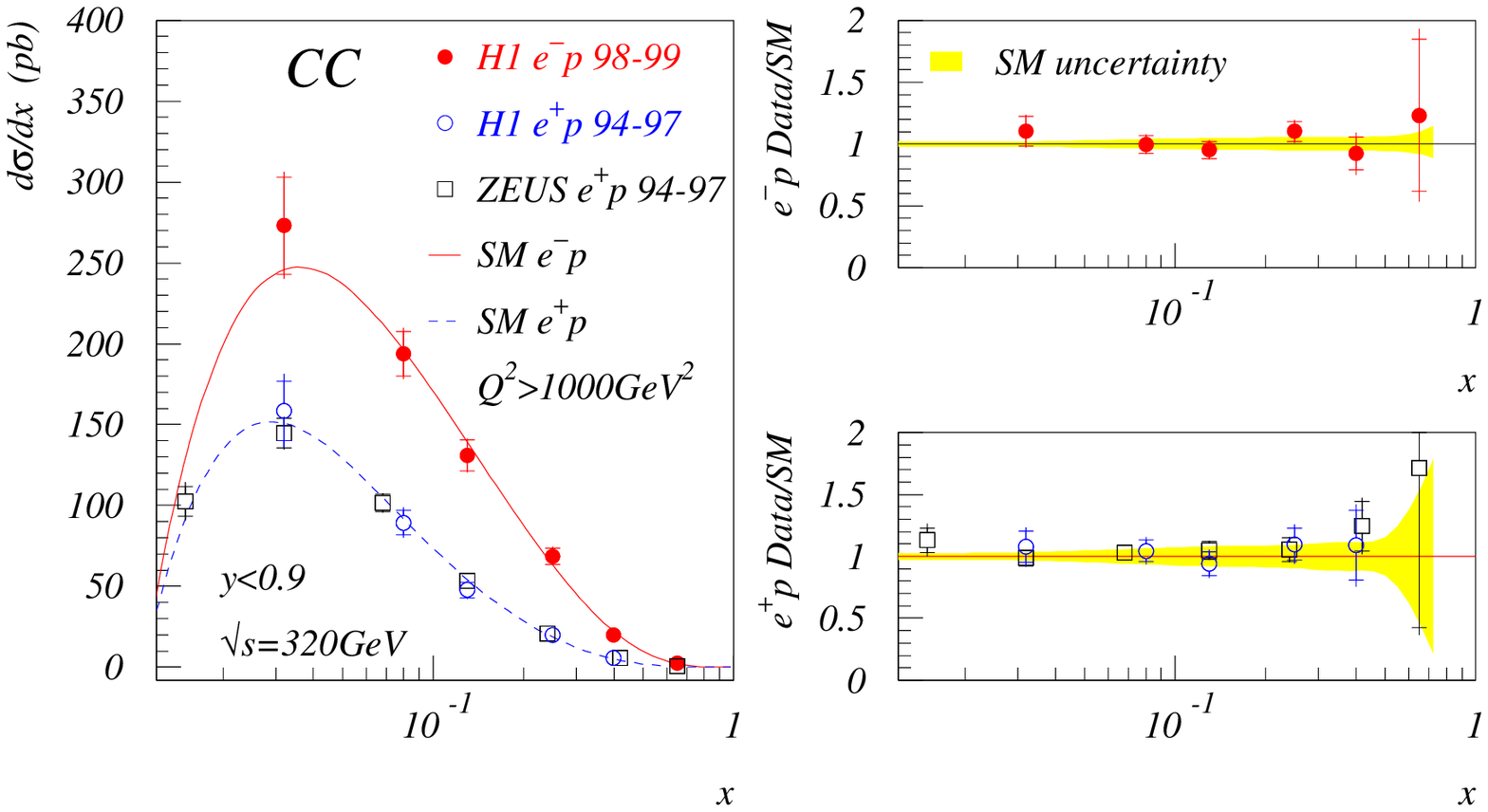,bbllx=0pt,bblly=0pt,bburx=594pt,
bbury=842pt,width=14cm}}
\end{picture}
\end{center}
\vspace*{0.5cm}
\caption{\label{fig:dsdx}The $x$ dependence of the measured NC (upper) and
CC (lower) cross-sections~\cite{h19899,h19497,h19900,zeus9497} for 
$Q^2>1000\,{\rm GeV}^2$ and $y<0.9$ 
in comparison with the Standard Model (SM) expectations, 
which are based on the H1 97 PDF fit~\cite{h19497}.}
\end{figure}

The $x$ dependence of the measured NC and CC
cross-sections~\cite{h19899,h19497,h19900} for
$Q^2>1000\,{\rm GeV}^2$ and $y<0.9$ is compared with the standard DIS
expectation in figure~\ref{fig:dsdx}. From the ratio plots, the NC
cross-sections at $x=0.65$ are seen to lie considerably below the expectations,
whereas the CC $e^+p$ cross-sections at large $x$ (dominated by the $d$
valence quark contribution) have the tendency to lie above the expectation
although the uncertainty of both the measurement and the expectation are large.

In order to quantify the impact of these measurements on the PDFs at large
$x$, two methods are employed to extract the $u$ and $d$ valence quark
densities using the HERA data alone.

\begin{figure}[h]
\vspace*{-3.5cm}
\begin{center}
\begin{picture}(50,370)
\put(-145,-125){\epsfig{figure=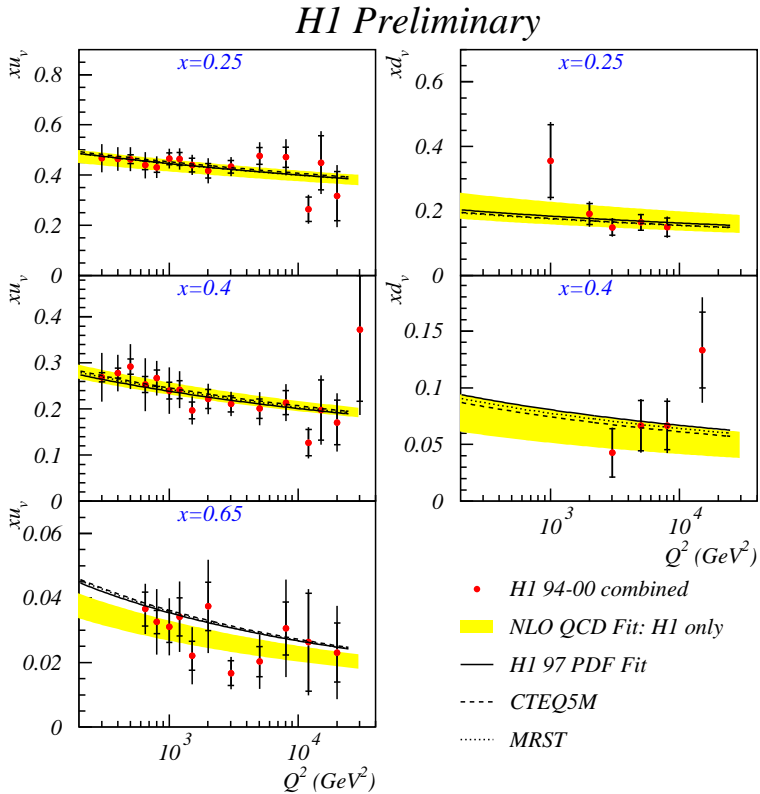,bbllx=0pt,bblly=0pt,bburx=594pt,
bbury=842pt,width=12.cm}}
\end{picture}
\end{center}
\vspace*{0.9cm}
\caption{\label{fig:h1udv}The valence quark densities $xu_v$ and $xd_v$
determined both with an NLO QCD fit using the H1 data only (shaded error
bands) and with a local extraction method (data points with the inner and full
error bars showing, respectively, the statistical and total uncertainties).
For comparison, three other parameterizations (H1 97 PDF Fit~\cite{h19497}, 
CTEQ5M~\cite{cteq} and MRST~\cite{mrst}) are also shown.}
\end{figure}

The first method in essence is a global NLO QCD fit like those performed by
the MRST and CTEQ groups~\cite{mrst,cteq}. The main difference is 
in the number of experimental data sets used.
The results of the H1 fit~\cite{zhang} 
are shown in figure~\ref{fig:h1udv} together with
those obtained from a second method. In the second method, 
the $u$ and $d$ quark densities are extracted locally
from the measured cross-sections $(\rmd^2\sigma/\rmd x\rmd Q^2)_{\rm meas}$ as:
\begin{equation}
xq_v(x,Q^2)=(\rmd^2\sigma(x,Q^2)/\rmd x\rmd Q^2)_{\rm meas}
\left(\frac{xq_v(x,Q^2)}{\rmd^2\sigma(x,Q^2)/\rmd x\rmd Q^2}\right)_{\rm th}
\end{equation}
where the second factor on the right-hand-side of the equation is the
theoretical expectation. Only those points where the $xq_v$ contribution is
greater than 70\% of the corresponding cross-section are
considered\footnote{The extracted parton densities are thus rather
independent of the theoretical input as the uncertainty on the dominant
valence quark contribution and that of the cross-section largely cancel in
the ratio.}. The first such extraction was performed by H1 for two values of
$x$ at 0.25 and 0.4 using the 1994-1997 $e^+p$ NC and CC
cross-sections~\cite{h19497}.
With the new $e^-p$ 1998-1999 and $e^+p$ 1999-2000 data, similar extractions
were made and were extended to $x=0.65$ for $u$~\cite{zhang}. In practice, 
the $d$ valence quark density is determined from the combined $e^+p$ CC
cross-sections, whereas the $u$ valence quark density is determined
from the combined $e^+p$ NC, $e^-p$ NC and $e^-p$ CC
cross-sections. Three independent determinations of $xu_v$ are then combined.

The valence quark distributions at large $x$ can thus be quantitatively 
constrained for the first time by the HERA data alone although 
the experimental uncertainties (between 6\% at $x=0.25, 0.4$ and $\sim 10$\% 
at $x=0.65$ for $xu_v$ and $\sim 20$\% for $xd_v$) are still large. 
The determined parton densities agree well with those parameterizations,
which use the fixed-target data as the main constraining source,
except for $xu_v$ at $x=0.65$, where the former is about $\sim 17$\% lower 
than the latter with little dependence on $Q^2$ in the covered kinematic 
range. The difference (less than two standard deviations) remains however 
not very significant.

\begin{figure}[h]
\begin{center}
\vspace*{1cm}
\begin{picture}(50,490)
\put(-125,220){
\hspace*{-1cm}\epsfig{figure=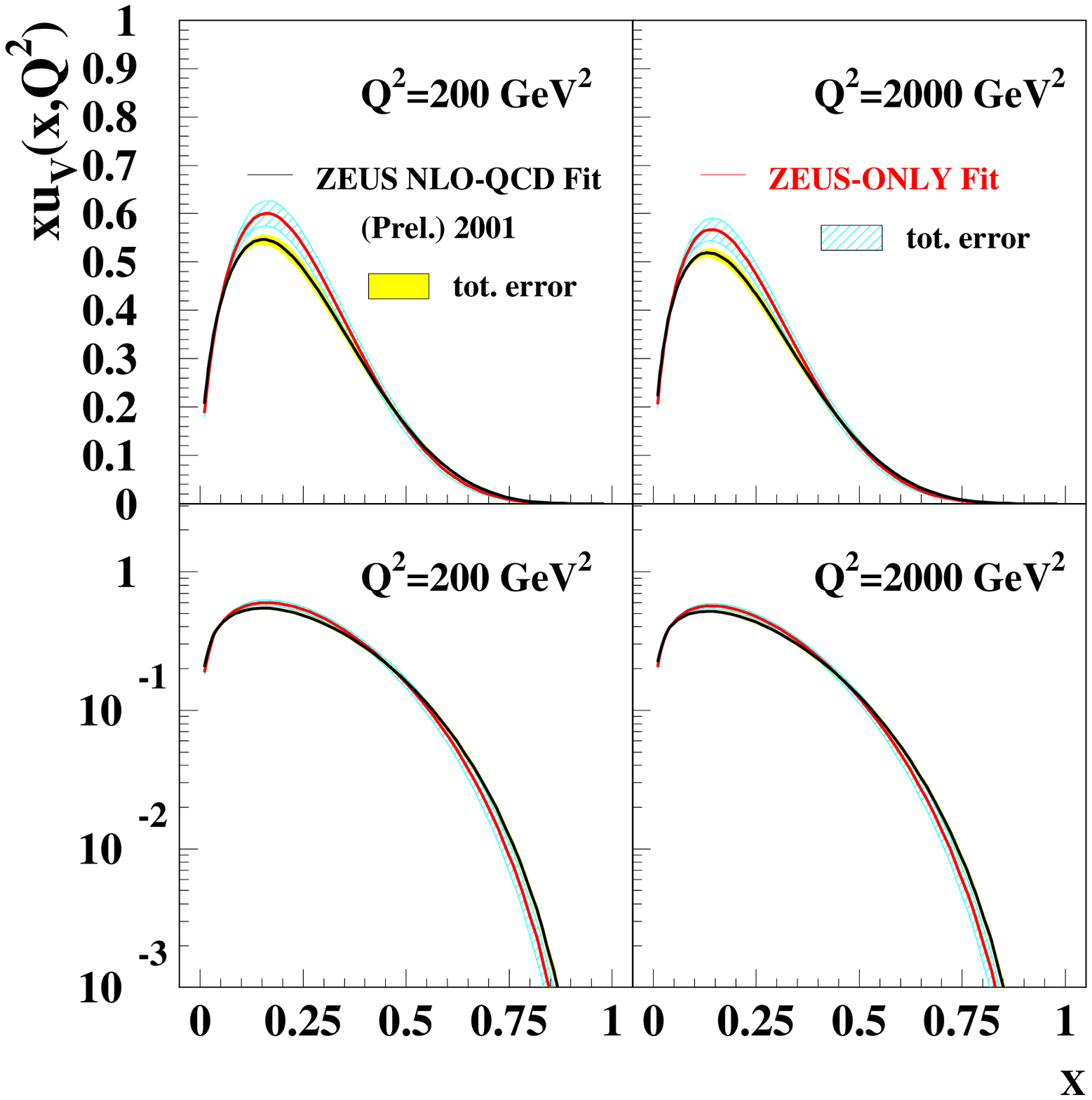,bbllx=0pt,bblly=0pt,bburx=594pt,
bbury=842pt,width=11.5cm,height=12.5cm}}
\put(-125,-0){
\hspace*{-1cm}\epsfig{figure=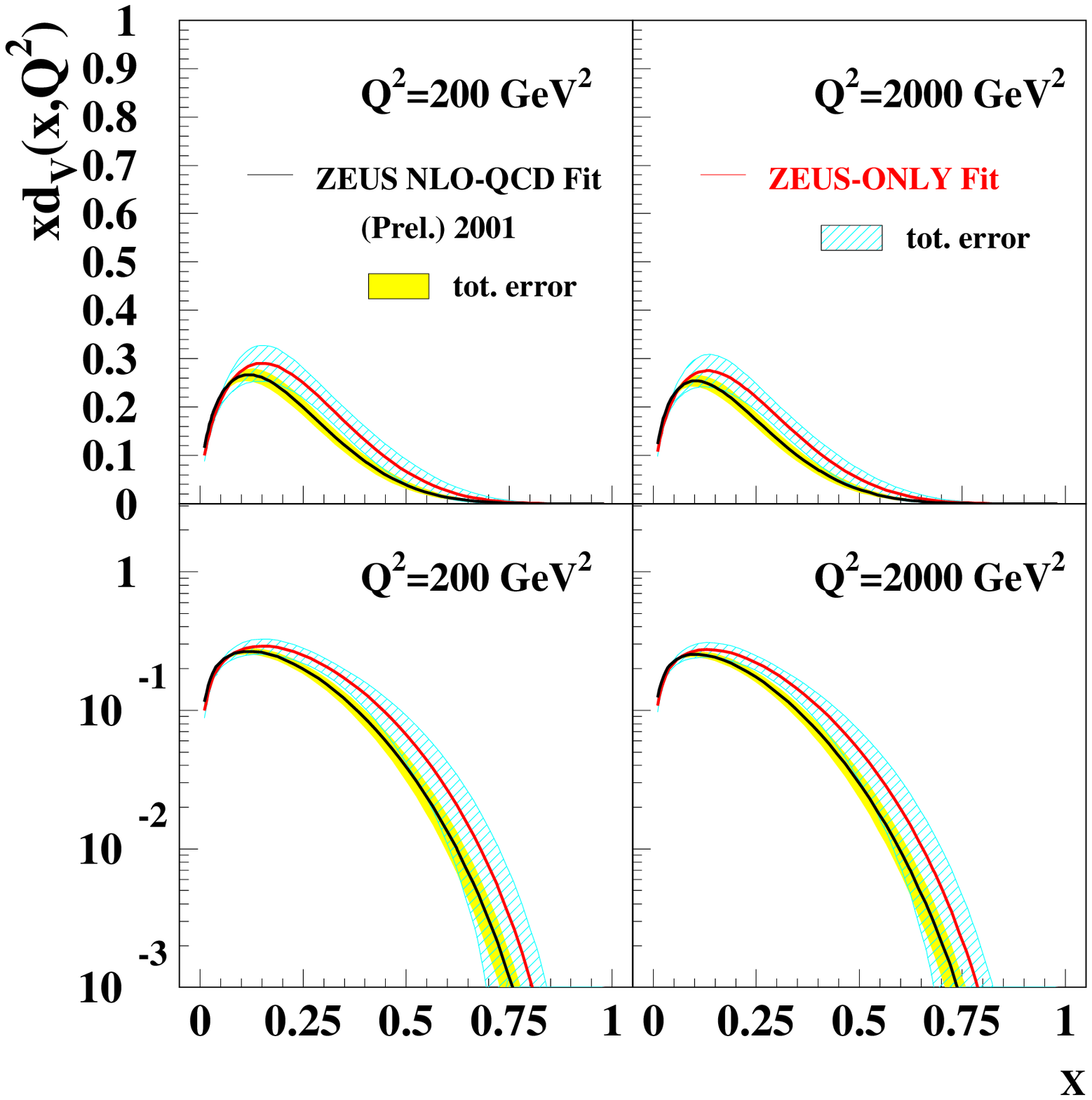,bbllx=0pt,bblly=0pt,bburx=594pt,
bbury=842pt,width=11.5cm,height=12.5cm}}
\end{picture}
\end{center}
\vspace*{-2cm}
\caption{\label{fig:zeusudv}The valence quark densities $xu_v$ (upper) and
$xd_v$ (lower) determined~\cite{zeusfit} for $Q^2=200\,{\rm GeV}^2$ and 
$2000\,{\rm GeV}^2$ in a fit with ZEUS data only in comparison with those 
determined in a global fit (ZEUS NOL-QCD Fit)~\cite{zeusfit} which uses both 
the fixed-target data and the 1996-1997 $e^+p$ ZEUS data.}
\end{figure}

Similar results~\cite{zeusfit} from ZEUS based on an NLO QCD fit using 
ZEUS data only are shown in figure~\ref{fig:zeusudv}. 
In comparison with the global QCD fit
which uses both the fixed-target data and the 1996-1997 $e^+p$ ZEUS data, the
ZEUS data prefer a larger $xu_v$ for $x$ around 0.2 and smaller $xu_v$ at
larger $x$. A shift towards large $x$ is also observed in the $xd_v$
distribution although the shift stays essentially within the uncertainty.

\section{Summary and future prospects} \label{sec:summary}
The structure function and other measurements from fixed-target DIS and
hadron-hadron collider experiments have provided us important inputs for
constraining the parton density distributions.
In the past few years there has been a renewed interest in the parton
density distributions at large $x$, in particular the behavior of the $d/u$
ratio when $x\rightarrow 1$. Considerable progress has been made towards
understanding some of the uncertainties in the individual measurements that
contribute to our knowledge of the large-$x$ parton distributions. The
current situation is that the large-$x$ distributions are less well constrained
than the medium-$x$ ones and need additional inputs for improvements.

HERA has made steady progress since 1992. The early runs have provided
unique structure function data for settling the behavior of parton (in
particular the gluon) densities at small $x$. The high statistics samples
taken in the recent years now allow the inclusive cross-sections be measured
for both neutral and charged current interactions at high $Q^2$. These
cross-sections have started to give quantitative constraints on the valence
quark densities at large $x$. HERA is finishing its upgrade program. After
the upgrade, the machine will provide about a factor of
five increase in the peak luminosity and polarized lepton beams. 
The upgraded machine and the improved detectors will thus
significantly improve in the next few years the measurement of 
the cross-sections and the knowledge of the parton densities at large $x$.
These data are unique as they are free from any nuclear corrections inherent
in the structure function data of the deuteron.

There are other possibilities by which the $u$ and $d$ valence quark
densities at large $x$ can be further constrained. One
possibility~\cite{melnitchouk98} is to use semi-inclusive DIS data on hadron
production in the current fragmentation region to measure the relative
yields of $\pi^+$ and $\pi^-$ mesons. The idea is fairly simple: at large
$z$ ($z$ being the fractional energy of the hadron), the $u$ quark
fragments primarily into a $\pi^+$, while a $d$ fragments into a $\pi^-$, so
that at large $x$ and $z$ one could have a direct measure of the $d/u$ ratio,
again free from the nuclear corrections when a proton target is used.

On the large-$x$ gluon density, future improvements are expected from the
direct photon data once the current discrepancies between the data and
the predictions and among the data are resolved. The improved Tevatron jet data
at Run\,II should also help. A third possibility~\cite{berger} is to use the
Drell-Yan process in the phase space where the leading-order subprocess
$qg\rightarrow \gamma^\ast q, \gamma^\ast\rightarrow \mu^+\mu^-$ dominates.

The uncertainties of the parton density distributions translate directly into
uncertainties in essentially every measurement made at a hadron-hadron
collider; it is therefore imperative that these distributions be well
determined.

\section*{Acknowledgments}
The author wishes to thank his colleagues in the H1 and ZEUS collaborations
for the measurements presented in this paper. He also thanks the organizers
of the workshop for the invitation and for the unique hospitality at the 
Ringberg castle.

\end{document}